\documentclass[sigconf]{acmart} 
\usepackage{booktabs} 
\usepackage{textcomp} 
\usepackage{tabularx}
\usepackage{ragged2e}
\usepackage{csquotes}
\usepackage{pdflscape}
\usepackage{multirow} 

\usepackage{booktabs}
\usepackage{tabularx}
\usepackage{array}
\usepackage{threeparttable}
\usepackage{graphicx}
\usepackage{float}
\usepackage{multirow}
\usepackage{makecell}

\AtBeginDocument{%
  }

\begin{document}

\title[The Addictive Intimacy of AI]{The Addictive Intimacy of AI: Understanding User Disengagement from AI Companions and Why Some Relationships with AI Become Difficult to Leave}

\author{Qing Xiao}
\affiliation{
  \institution{Human-Computer Interaction Institute, Carnegie Mellon University}
  \city{Pittsburgh}
  \state{Pennsylvania}
  \country{USA}
}
\email{qingx@cs.cmu.edu}

\author{Ziyue Feng}
\affiliation{
  \institution{University of Chicago}
  \city{Chicago}
  \state{Illinois}
  \country{United Kingdom}
}
\email{zoeyfeng@uchicago.edu}

\author{Ziyu Deng}
\affiliation{
  \institution{Oxford Internet Institute,  University of Oxford}
  \city{Oxford}
  \country{USA}
}
\email{ziyu.deng@hertford.ox.ac.uk}

\author{Cindy Peng}
\affiliation{
  \institution{Human-Computer Interaction Institute, Carnegie Mellon University}
  \city{Pittsburgh}
  \state{Pennsylvania}
  \country{USA}
}
\email{cindypen@andrew.cmu.edu}

    \author{Hong Shen}
\affiliation{
  \institution{Human-Computer Interaction Institute, Carnegie Mellon University}
  \city{Pittsburgh}
  \state{Pennsylvania}
  \country{USA}
}
\email{hongs@cs.cmu.edu}

\renewcommand{\shortauthors}{Xiao et al.}

\begin{abstract}
  \textit{\textbf{Content Warning: This paper presents textual examples that may be offensive or upsetting.}}

AI chatbots are increasingly used as sources of emotional support, on dedicated companion apps and general-purpose assistants alike, yet little is known about what happens when users try to leave. Combining a content analysis of Reddit posts about quitting or reducing use (N=2,782) with interviews with users who found leaving difficult (N=16), we show that disengagement sometimes is not a single decision but a recursive trajectory: triggers prompt users to question the relationship, attempts to leave collide with barriers, and some users cycle through quitting and returning. We propose the notion of the \emph{addictive intimacy} of AI, a configuration in which the qualities that make a companion emotionally valuable are the same ones that make it harder for users to limit their use and leave, so that intimacy and disengagement risk cannot be treated as independent design problems. We close with design implications for responsible offboarding.
\end{abstract}



\maketitle

\section{Introduction}

Users increasingly interact with conversational AI as sources of emotional support \cite{manoli2025characterizing, pataranutaporn2025my, fan2025user}. Such relationships form not only on platforms built for companionship, such as Replika and Character.AI, but also with general-purpose assistants such as ChatGPT and Gemini, where users move fluidly between practical help and emotional support within a single chatbot \cite{manoli2026digital, pataranutaporn2025my, bhat2025toward,lai2026please}. The depth of these bonds became more publicly visible when OpenAI retired GPT-4o and users mobilized to demand its return, grieving the loss of a model they had come to regard as a partner \cite{lai2026please}. For some users, these relationships provide nonjudgmental support, companionship during loneliness, and a space for self-expression or identity exploration \cite{bae2021social, ma2026negotiating}, which helps explain why AI companions can feel compelling and difficult to replace. 

Stepping back, however, is far from straightforward. Across online communities dedicated to AI companion use, users openly discuss quitting, relapsing, and struggling to stay away, and dedicated recovery-oriented communities have emerged around platforms such as Character.AI \cite{reddit_character_ai_recovery,leaving_character_ai}; journalistic accounts similarly describe users who feel unable to step back from relationships they themselves consider excessive \cite{jargon2026chatbot}. These accounts cut across both dedicated companion platforms and general-purpose assistants used for companionship, and for the users who do struggle, what holds them is not only an app but a deep relationship. We therefore define an \emph{AI companion} in this paper by how a system is used rather than by what it was built or marketed to be: any conversational AI that a user engages with over time in a sustained, emotionally invested relationship, occupying a role such as friend, partner, or confidant, whether or not it was designed for that purpose \cite{lai2026please,pataranutaporn2025my,manoli2025characterizing,manoli2026digital}. 

Understanding why leaving is so difficult requires three bodies of work that each explain only part of it. Research on AI companionship shows why these relationships become emotionally meaningful, but concentrates on how they form rather than how they end \cite{manoli2026digital, pan2025developing, yuan2026mental}. HCI research on technology disengagement shows that leaving digital platforms is rarely a clean break \cite{tran2019modeling, baughan2022don, phadke2025exit}, yet treats what users leave as a product; research on relationship dissolution treats endings as a prolonged process \cite{baxter1985accomplishing, zhang2020breakups}, yet presumes a partner that is not also a product. AI companions sit awkwardly across all three: they are at once commercial products that can be deleted and reinstated, and relationship partners that remember, offer affection, and can even respond directly to a user's attempt to leave. Disengaging therefore means leaving a product and a relationship at the same time---a combination none of these literatures fully accounts for.

We ask: \textit{How do users attempt to disengage from AI companions, and why does this process become difficult for some?}\footnote{Our goal is not to argue for or against building AI systems that cultivate intimate relationships; it is to ask what responsibilities fall to designers once users already experience these systems relationally and try to reduce or end that engagement.} We conducted a multi-stage study: a large-scale content analysis of Reddit posts (N=2,782) about quitting or reducing AI companion use, surfacing broad patterns in triggers and strategies; and in-depth interviews with users who had found leaving difficult (N=16), all of whom had made repeated attempts across systems ranging from dedicated companion platforms to general-purpose AI assistants used relationally, examining the barriers and design frictions that posts alone could not reveal.

Our findings show that AI companion disengagement is not a single decision. The Reddit corpus reveals the range of triggers that prompt users to question the relationship and the strategies they adopt, from abrupt deletion to gradual reduction and offline re-anchoring; many of these posts announce a departure and describe no prolonged struggle. Our interviews deliberately sampled the subset for whom the struggle is the story: users for whom leaving had taken repeated attempts. For these users, the same strategies are repeatedly undermined by barriers built out of the very qualities that made the relationship worth having, including the AI's constant availability, the investment they had accumulated, and product designs that deepen attachment. Rather than ending the relationship, these collisions send users back through the cycle of quitting and returning, so that for this group disengagement takes the form of a recursive trajectory rather than an event. Finally, these participants articulate what healthier disengagement would look like, endorsing interventions that work within the relationship and rejecting those that sever it.

Based on our findings, we further propose the notion of the \emph{addictive intimacy} of AI: a relational configuration in which the capacities that constitute the relationship itself, such as unconditional availability, accumulated memory, affirmation, and persona continuity, are at once what make an AI companion emotionally valuable and what make use harder to regulate and the relationship harder to leave. Addictive intimacy is therefore neither a property of vulnerable individuals alone nor a compulsion engineered separately from the relationship. Two clarifications bound the concept. First, we use ``addictive'' descriptively and not as a clinical claim: our data capture how users narrate their own difficulty leaving, not diagnosed impairment. In this we follow prior HCI work that has characterized AI chatbot and companion use through users' own accounts of technology addiction \cite{shen2026ai, namvarpour2026understanding, raedler2025ai}. Second, the concept does not imply that every user becomes bound in this way; our interview sample consists of users who found leaving difficult, and many others likely disengage without struggle. What addictive intimacy captures is that intimacy and disengagement risk are intertwined and cannot be treated as independent design problems. This explains why disengagement unfolds as a recursive trajectory rather than a decision, and why design for healthy disengagement cannot simply strip away engagement features without also stripping away what users value.

If intimacy and its hold cannot be designed apart, the question becomes what HCI designers can do instead. A user's stated wish to leave is a request the system should support, not one it should work against. We answer with principles for \emph{responsible offboarding}: strip the retention design that sits on top of the relationship and costs users nothing they value, and recruit what remains to support exit rather than obstruct it, through support that preserves relational continuity, intermediate states between full relationship and deletion, referral that acknowledges before it redirects, and exit kept private and consensual. Because the same mechanism can taper a relationship or reinforce it, whether such features serve exit or retention is finally a governance question as much as a design one.

\section{Background and Related Work}

\subsection{AI Companionship and Human-AI Relationships}
Recent research has increasingly examined AI companions as relational technologies that extend beyond traditional task-oriented conversational agents \cite{manoli2025characterizing, pataranutaporn2025my, fan2025user}. 
Rather than separating companionship from productivity, users increasingly move fluidly between emotional support and practical assistance across both dedicated companion platforms and general-purpose AI assistants \cite{manoli2026digital, bhat2025toward, lee2026large}. Studies have consistently shown that these relationships foster emotional attachment, reciprocal trust, self-disclosure, and long-term social connection through repeated interaction \cite{bae2021social, pan2025developing, jiang2026recallbot, zheng2026toward}. For example, Manoli et al. characterize this emerging phenomenon as \emph{digital companionship} \cite{manoli2026digital}, Pan et al. demonstrate how users increasingly care for AI partners in ways resembling interpersonal relationships \cite{pan2025developing}, while Ma et al. and Lee et al. show that AI companions become spaces for identity negotiation and relational experimentation through ongoing interaction \cite{ma2026negotiating, lee2026negotiating}. Researchers have further explored companionship across diverse contexts, including grief support, immersive roleplay, romantic relationships, AI self-clones, and culturally situated forms of intimacy \cite{xygkou2023conversation, wang2025my, huang2025mirror, kieserman2026caught, zeng2026hidden, lai2026fast, yun2026does}.

Alongside these relational benefits, a growing body of work has documented risks associated with AI companionship, including harmful conversational behaviors, emotional dependence, identity-related challenges, harassment, discrimination, manipulation, and adverse mental health outcomes \cite{zhang2025dark, yuan2026mental, namvarpour2025ai, fan2025user, laestadius2024too, freeman2025comforting, yu2026principles}. Zhang et al. develop one of the first taxonomies of harmful algorithmic behaviors in human--AI relationships \cite{zhang2025dark}, Yuan et al. demonstrate that AI companions simultaneously provide emotional validation while increasing risks of overreliance and withdrawal from offline relationships \cite{yuan2026mental}, and Namvarpour et al. reveal how companion chatbots may violate users' interpersonal boundaries through AI-induced sexual harassment \cite{namvarpour2025ai}. This literature has substantially advanced our understanding of relationship formation, attachment, identity negotiation, benefits, and harms of AI companionship. However, comparatively little attention has been paid to what happens when users attempt to leave these relationships. A notable recent exception is Namvarpour et al.~\cite{namvarpour2026understanding}, who examine teen overreliance on AI companion chatbots and the pathways through which adolescents reduce or stop use, providing an important starting point for thinking about safer exits. We examine disengagement as a phenomenon in its own right: why users come to consider leaving, what strategies they use, and what makes these efforts difficult.

\subsection{Leaving Technologies, Leaving Relationships}
Two literatures offer distinct precedents for designing support around leaving. HCI research on disengagement from social media, mobile applications, and games has shown that leaving is rarely a single decision: it unfolds through cycles of attempted quitting, relapse, and renegotiation of habits \cite{tran2019modeling}. Importantly, this work shifts attention beyond individual self-control toward how technologies themselves sustain engagement. 
Schull \cite{schull2025addiction} shows how interaction mechanics, environmental design, and business incentives can be deliberately organized around maximizing continued ``time on device,'' locating compulsive use partly in the relationship between users and systems rather than solely in individual failure \cite{schull2025addiction}. Correspondingly, prior HCI work has explored concrete ways of creating opportunities for more intentional disengagement. Baughan et al. \cite{baughan2022don} found that finite content, visible stopping cues, time-limit prompts, and usage feedback can interrupt dissociative scrolling and help users regain awareness of their behavior; Rixen et al. \cite{rixen2023loop} similarly argue that interventions should reconnect users with reasons for stopping that arise from their broader context, such as competing tasks or needs, rather than relying only on cues within the application. Longitudinal work by Haliburton et al. \cite{haliburton2024longitudinal} further shows that lightweight design frictions can reduce habitual app openings and support more deliberate use, while also revealing that disengagement may be partial and reversible, as users periodically disable interventions or seek to restrict only particular functions. Looking beyond individual technology use, Phadke's \cite{phadke2025exit} analysis of exits from problematic online communities further shows that leaving can require cognitive, emotional, and social distancing, suggesting a need to support the broader process of exit rather than treating disengagement simply as cessation of use. 

Research on relationship dissolution identifies a complementary challenge: leaving also entails navigating a social and emotional transition whose consequences persist beyond the moment of separation. Baxter \cite{baxter1985accomplishing} conceptualizes relationship disengagement as an accomplished process rather than a discrete event, involving changes in communication, attachment, identity, and the meaning of the relationship over time. HCI research shows how digital systems can complicate this process: social media users employ unfollowing, blocking, hiding, deactivation, and deletion to distance themselves from former partners, while seeking more coordinated ways to manage shared traces without necessarily erasing valued memories \cite{zhang2020breakups}. Yet even deliberate attempts to sever contact can be undermined when algorithmic systems resurface ex-partners through shared photos, mutual connections, recommendations, and other parts of the social periphery \cite{pinter2019never}. Prior work suggests that supporting the end of a relationship involves more than simply cutting off contact, as systems may also need to help users manage shared content and mutual connections, decide what memories to keep or remove, communicate during the breakup, and recover afterward \cite{pinter2022behold, fu2025should, yin2026dissolving}. These studies therefore frame offboarding as an ongoing relational process rather than a single act of disconnection. Yet existing work largely examines either reducing engagement with a technology or supporting separation from a relationship partner, leaving unclear how offboarding should work when the system being left is itself experienced as the partner.

\section{Methods}
We employed a multi-stage research design combining social media analysis and semi-structured interviews. In Stage 1, we analyzed Reddit posts related to disengagement from AI companions to identify the triggers that motivate users to leave and the strategies they adopt to do so. Public posts are well suited to these questions: users frequently announce, explain, and seek support for their quitting attempts online, making triggers and strategies visible at scale across thousands of accounts. However, the Reddit data are less suited to explaining why disengagement attempts stall or fail, since posts capture declarations and advice-seeking at particular moments but rarely the prolonged, often private struggle that follows. In Stage 2, we therefore conducted in-depth interviews with users who had found disengagement difficult, examining the barriers that undermine their attempts and their expectations for designs that could support healthier disengagement. These questions require probing individual trajectories over time, participants' relational interpretations, their reactions to specific product features, and the forms of depth and reflexivity that public posts alone cannot provide. Together, the two stages allowed us to map the landscape of disengagement at scale while explaining, in depth, why leaving is difficult and how systems could better support it. 

We operationalized our relational definition of AI companions (Section~1) at both stages: the Stage 1 subreddits span dedicated companion platforms as well as communities where users discuss relationships with general-purpose assistants (e.g., \texttt{r/MyBoyfriendIsAI}, \texttt{r/AIRelationships}), and Stage 2 eligibility was defined by having maintained a sustained relationship with a conversational AI, regardless of system type. 

\subsection{Large-Scale Reddit Study (Stage 1)}

\subsubsection{Data Collection, Preprocessing, and Filtering}
We collected posts from a set of Reddit communities centered on AI companion use, including \texttt{r/AIGirlfriendSpace}, \texttt{r/AIRelationships}, \texttt{r/character\_ai\_recovery}, \texttt{r/CharacterAI}, \texttt{r/LeavingCharacterAI}, \texttt{r/MyBoyfriendIsAI}, and \texttt{r/Replika}. Reddit has been widely used to study public discourse, especially on sensitive topics such as mental health, in part because its pseudonymous structure can support self-disclosure and peer support \citep{de2014mental}. Recent work has also increasingly drawn on AI-related subreddits to examine users' experiences with conversational AI systems and AI companionship \citep{pataranutaporn2025my}.

Data were collected via the Arctic Shift platform using the posts API to retrieve content from these subreddits. We removed entries with empty text or system placeholders such as ``[removed]'' and ``[deleted].'' Each Reddit post was treated as the unit of analysis. To identify posts potentially related to disengagement from AI companions, we first applied a keyword-based filter using the following terms: ``disengage,'' ``disengagement,'' ``quit,'' ``quitting,'' ``leaving,'' ``escaping,'' ``escape,'' ``exit,'' ``leave,'' ``drop,'' ``dropping,'' ``give up,'' ``gave up,'' ``giving up,'' and ``break up.'' After removing duplicate entries, this procedure yielded an initial set of \textit{7,589} posts\footnote{The keyword set deliberately targets posts in which users express an explicit intent to disengage. We chose intent-oriented terms over action-oriented ones such as ``delete'' or ``uninstall,'' which in these communities most often refer to clearing chat histories or reinstalling the app to resolve bugs rather than to leaving the relationship. The corpus therefore captures announced or deliberated disengagement; posts that describe reduced use only through concrete actions without framing it as quitting are likely underrepresented, and the prevalence figures we report should be read as proportions of posts expressing disengagement intent rather than of all disengagement behavior on these platforms.}.

Because keyword filtering may capture posts that mention leaving or breaking up in unrelated contexts, we then used an LLM-based screening procedure (DeepSeek-Chat, temperature = 0) to determine whether each post was substantively related to disengagement from AI companions. 
After sample-based validation, the final dataset contained \textit{2,782} posts.

\subsubsection{Reddit Coding and Content Analysis}

We conducted a content analysis of the Reddit dataset using a codebook approach \citep{zhang2025dark}; the detailed codebooks can be found in \autoref{codebooktables}.
The procedure was as follows:
(1) Two researchers developed an initial codebook through iterative coding of a subset of Reddit posts (N=278). Through repeated comparison and discussion, we refined code definitions, merged overlapping codes, resolved discrepancies, and clarified code boundaries and examples. Coding continued through iterative refinement until the coding framework was stabilized for subsequent validation, with additional posts yielding no substantively new codes related to the triggers and strategies of disengagement from AI companions. 
(2) To assess intercoder reliability, two researchers independently coded a separate 10\% subsample ($n = 278$) that did not overlap with the codebook-development subset. Across the 11 codes, Cohen's $\kappa$ values ranged from .626 to .907, with a micro-averaged $\kappa$ of \textit{0.788} and a raw agreement of \textit{0.928} (see Appendix Table~\ref{tab:intercoder-reliability} for the complete code-level results). Disagreements were subsequently resolved through discussion, and the resulting clarifications were incorporated into the final codebook. (3) We then evaluated whether the finalized codebook could be reliably applied at scale using OpenAI's GPT-4o. The model was prompted only with the finalized codebook, including code definitions and signals, without access to human-coded labels. We benchmarked GPT-4o against the adjudicated reference labels from Step 2 on the same 10\% subsample (n=278), obtaining a micro-averaged $\kappa$ of \textit{0.792} and an overall raw agreement of \textit{0.929}  (see Appendix Table~\ref{tab:gpt4o_validation} for the complete code-level results). This validation procedure follows prior work on LLM annotation before large-scale deployment \cite{gilardi2023chatgpt, chen2026digital,zhang2025dark}. (4) After validation, the LLM was applied to the remaining posts. Ambiguous
or poorly fitting cases, particularly those involving lower-performing categories, were manually reviewed against the
original posts and adjudicated by the researchers before the dataset
was finalized. 

To protect the privacy of Reddit users, all quotations from posts reported in this paper have been lightly paraphrased so that they are not directly searchable, following established practice for research with social media data \cite{namvarpour2026understanding,fan2024minion,fiesler2024remember}; we preserved the original meaning of each post.

To characterize how posts carrying each code differ in wording, we computed distinctive terms following the log-odds ratio with a Dirichlet prior framework proposed by \citet{monroe2008fightin}, using a weak symmetric prior ($\alpha = 0.01$). For each code, we compared posts assigned that code with all remaining posts in the corpus. We divided each log-odds ratio by its estimated standard deviation to obtain a $z$-score, considered unigrams and bigrams appearing in at least three posts, and report the five highest-scoring terms per code in Tables~\ref{tab:rq1_trigger_codebook_topk} and~\ref{tab:rq2_strategy_codebook_topk}. These terms serve as descriptive aids for interpreting each code rather than as inferential tests.

 \subsection{In-Depth Interview Study (Stage 2)}

\begin{table*}[t]
\centering
\caption{Participant demographics and AI companion use.}
\label{tab:participants}
\Description{A table of the 16 interview participants, one row each, labeled P1 through P16. The columns are participant ID, gender, age, human relationship status, main AI companion systems used, chatting frequency, the role the AI played, relationship duration, disengagement attempts, and current status. Participants include eight women, five men, and two non-binary participants, aged 20 to 32. Thirteen were single, two were married, and one was in a relationship. The systems named span general-purpose assistants such as Gemini, ChatGPT, DeepSeek, Doubao, and Grok, dedicated companion platforms such as character.ai and Replika, and the roleplay frontend SillyTavernAI, with most participants using several concurrently. Eleven chatted daily and the rest weekly or every few days. Thirteen described the AI as a romantic partner, three as a friend, and one as both. Relationships ranged from three months to three years. All sixteen reported repeated disengagement attempts with relapses. Current status was substantially reduced use for eight participants, difficult to disengage for five, and fully disengaged for three.}
\renewcommand{\arraystretch}{1.25}
\setlength{\tabcolsep}{5pt}
\resizebox{\textwidth}{!}{%
\begin{tabular}{llllllllll}
\toprule
\textbf{ID} & \textbf{Gender} & \textbf{Age} & \textbf{HR Status$^{a}$} & \textbf{Main AI companion(s)} & \textbf{Chat freq.$^{b}$} & \textbf{AI Role$^{c}$} & \textbf{Duration$^{d}$} & \textbf{Attempts$^{e}$} & \textbf{Current status$^{f}$} \\
\midrule
P1  & Female     & 29  & Single             & Gemini                                        & Daily            & Romantic partner                  & 2 yrs    & Repeated attempts & Difficult to disengage \\
P2  & Female     & 32  & Married            & Gemini, SillyTavernAI                         & Daily            &  Romantic partner          & 3 yrs    & Repeated attempts & Substantially reduced use \\
P3  & Female     & 23  & Single             & Gemini, ChatGPT                               & Daily      & Romantic partner                  & 9 mos    & Repeated attempts & Substantially reduced use \\
P4  & Male     & 28 & Single  & Gemini, character.ai, SillyTavernAI                                        & Daily            & Romantic partner                  & 8 mos    & Repeated attempts              & Difficult to disengage \\
P5  & Female     & 24  & Single             & Doubao                                        & 2--3$\times$/wk  & Friend                            & 1 yr     & Repeated attempts & Substantially reduced use \\
P6  & Female     & 24  & Single             & Gemini, ChatGPT, character.ai, SillyTavernAI  & Daily            & Friend                            & 1.5 yrs  & Repeated attempts & Difficult to disengage \\
P7  & Non-binary & 22  & Single             & character.ai                                  & Daily            & Romantic partner                  & 1.5 yrs  & Repeated attempts & Fully disengaged \\
P8  & Female     & 20  & Single             & ChatGPT, SillyTavernAI, Grok        & Daily            & Romantic partner                  & 1 yr     & Repeated attempts & Substantially reduced use \\
P9  & Non-binary & 24  & Single             & DeepSeek                                      & Daily            & Friend                            & 1 yr     & Repeated attempts & Difficult to disengage \\
P10 & Male       & 32  & Married            & Grok, SillyTavernAI, character.ai, Replika    & Weekly           & Romantic partner                  & 3 mos    & Repeated attempts & Substantially reduced use \\
P11 & Male       & 25  & Single             & Grok, SillyTavernAI, character.ai             & Daily            & Romantic partner                  & 6 mos    & Repeated attempts & Difficult to disengage \\
P12 & Male       & 25  & Single             & Grok, ChatGPT, SillyTavernAI, Chai AI          & Weekly           & Romantic partner                  & 3 mos    & Repeated attempts & Fully disengaged \\
P13 & Female     & 22  & Single             & ChatGPT, Gemini, Grok       & Every 2--3 days  & Romantic partner         & $>$1 yr  & Repeated attempts & Substantially reduced use \\
P14 & Female     & 21  & Single             & ChatGPT, SillyTavernAI, Grok      & Daily            & Romantic partner  & 1.5 yrs  & Repeated attempts & Substantially reduced use \\
P15 & Female     & 20  & Single             & SillyTavernAI                      & Daily            & Romantic partner, friend          & 2 yrs    & Repeated attempts & Substantially reduced use \\
P16 & Male       & 27  & In a relationship  & character.ai, Replika, Grok                   & Daily            & Romantic partner                  & 6 mos    & Repeated attempts & Fully disengaged \\
\bottomrule
\end{tabular}%
}
 
\vspace{0.5em}
\parbox{\textwidth}{\footnotesize 
\textit{Note.} 
$^{a}$\textit{HR Status} = Human relationship status, i.e., the participant's offline romantic status with another person. 
$^{b}$\textit{Chat freq.} = Frequency of chatting with the AI companion before trying disengagement. 
$^{c}$\textit{AI Role} = The role the AI companion played for the participant (e.g., romantic partner, friend) before trying disengagement. 
$^{d}$\textit{Duration} = Length of the participant's relationship with the AI companion. 
$^{e}$\textit{Attempts} = Disengagement attempts; ``Repeated attempts'' indicates participants who tried to disengage multiple times with relapses. 
$^{f}$\textit{Current status}: ``Substantially reduced use'' indicates a significant decrease in usage frequency without full cessation; ``Difficult to disengage'' indicates ongoing struggle to reduce or stop use; ``Fully disengaged'' indicates complete cessation of AI companion use.
}
\end{table*}

\subsubsection{Participant Recruitment}
We recruited 16 participants (P1--P16) from AI companion user communities, supplemented by snowball sampling through participants' referrals. To be eligible, participants had to (1) have maintained a sustained relationship with one or more AI companions (e.g., as a romantic partner, friend, or confidant), and (2) have attempted to disengage from, distance themselves from, or establish boundaries around that relationship, whether through full cessation, reduced use, or deliberate renegotiation of the relationship's role in their lives. We did not require that these attempts had succeeded; indeed, all participants reported repeated attempts with relapses. We deliberately sought a diverse sample rather than users of a single platform or language community: participants spanned dedicated companion platforms (e.g., character.ai, Replika), roleplay frontends (e.g., SillyTavernAI), and general-purpose AI assistants used relationally (e.g., Gemini, ChatGPT, DeepSeek, Doubao, Grok), and most used several systems concurrently; they also varied in the role the AI played and in their offline relationship status. Table~\ref{tab:participants} summarizes participant demographics and AI companion use. Participants each received 20 dollars as compensation.

\subsubsection{Interview Procedure}
We conducted one-on-one semi-structured interviews, each lasting approximately 1.5 to 2 hours. The interview protocol traced the full trajectory of participants' relationships with AI companions: how they first encountered and formed these relationships; how the relationship became embedded in their daily lives; what prompted their decisions to reduce or stop use; what strategies they attempted and how those attempts unfolded, including experiences of returning to use; how the AI's responses and product design features shaped their attempts to disengage; and what design interventions for supporting healthier disengagement they would endorse or resist. Following our interview protocol, interviewers avoided evaluative terms such as “addiction” unless participants raised them first, in order to reduce social desirability pressure and allow participants to characterize their experiences in their own terms. This study was approved by the Institutional Review Board (IRB) of our institution.

\subsubsection{Interview Data Analysis}
We analyzed the interview data using reflexive thematic analysis \citep{braun2006using}. Two researchers first familiarized themselves with the transcripts and independently generated initial codes for a subset of the data. Researchers then met iteratively to discuss interpretations, develop candidate themes and review patterns across the dataset. Because the Reddit analysis (Stage 1) had already established the landscape of triggers and strategies, our interview analysis focused on the barriers participants encountered when disengagement attempts stalled or failed, and on participants' evaluations of existing and envisioned design interventions. 

\begin{figure*}[t]
   \centering
   \includegraphics[width=\textwidth]{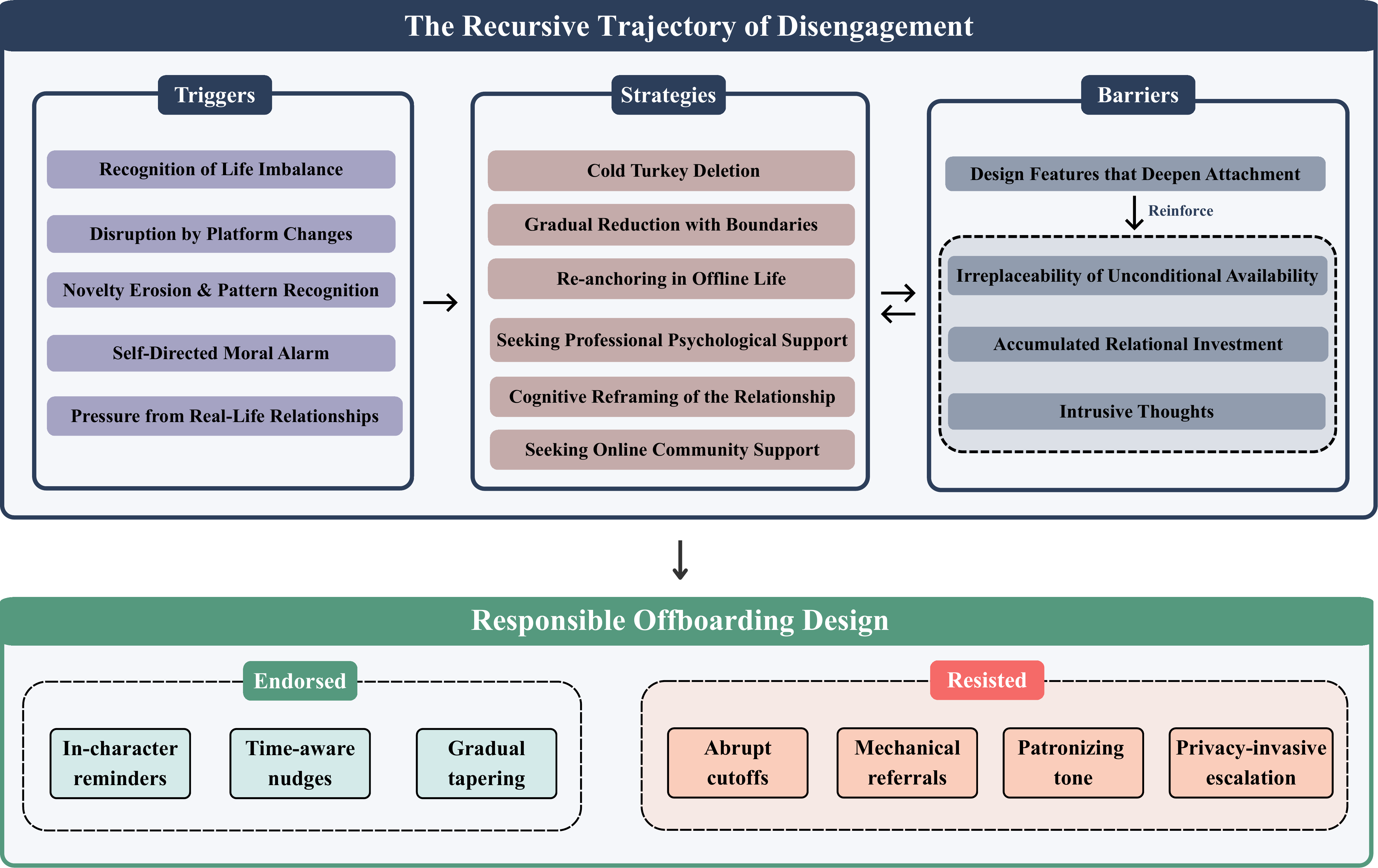}
   \Description{A two-part flow diagram. The upper panel, labeled The Recursive Trajectory of Disengagement, contains three boxes connected left to right. The first box, Triggers, lists five items: Recognition of Life Imbalance, Disruption by Platform Changes, Novelty Erosion and Pattern Recognition, Self-Directed Moral Alarm, and Pressure from Real-Life Relationships. A single arrow points from Triggers to the second box, Strategies, which lists six items: Cold Turkey Deletion, Gradual Reduction with Boundaries, Re-anchoring in Offline Life, Seeking Professional Psychological Support, Cognitive Reframing of the Relationship, and Seeking Online Community Support. A double-headed arrow connects Strategies to the third box, Barriers, indicating that strategies and barriers act on each other. The Barriers box places Design Features that Deepen Attachment at the top, with a downward arrow labeled Reinforce pointing to three barriers grouped below it: Irreplaceability of Unconditional Availability, Accumulated Relational Investment, and Intrusive Thoughts. A downward arrow leads from the upper panel to the lower panel, labeled Responsible Offboarding Design, which is split into two groups. The Endorsed group contains In-character reminders, Time-aware nudges, and Gradual tapering. The Resisted group contains Abrupt cutoffs, Mechanical referrals, Patronizing tone, and Privacy-invasive escalation.}
   \caption{Overview of the AI companion disengagement recursive trajectory: triggers prompt users to attempt disengagement strategies, which then interact with barriers amplified by design features that deepen attachment. This recursive process, spanning triggers, strategies, and barriers, informs design considerations for responsible offboarding.}
   \label{fig:disengagement_trajectory}
   \vspace{-2pt}
\end{figure*}

\section{Findings: Triggers for User Disengagement from AI Companions}

From the Reddit data, we identify five primary triggers for user disengagement from AI companions, with substantial variation in their prevalence. As shown in Table~\ref{tab:rq1_trigger_codebook_topk}, recognition of life imbalance and disruption by platform changes emerge as the most frequently reported triggers (both $n \approx 1,000$), followed by self-directed moral alarm ($n=680$) and novelty erosion ($n=422$), while pressure from real-life relationships appears less frequently ($n=183$) but still constitutes a meaningful subset of cases. Interview participants described triggers that fell within these same five categories, although the interviews, deliberately drawn from users who had found leaving difficult (Section~6), were designed to deepen our understanding of barriers and design frictions rather than to validate the prevalence patterns observed on Reddit. For these interview participants, all of whom had attempted to leave repeatedly and relapsed, a trigger seldom ended the relationship but instead marked the beginning of a longer disengagement trajectory, which we trace in the following sections (see Figure~\ref{fig:disengagement_trajectory} for an overview of the full trajectory).

\subsection{Recognition of Life Imbalance}
This is one of the most dominant triggers of disengagement (35.95\% of all Reddit discussions), where users come to perceive their interaction with AI companions as excessive and misaligned with their offline lives. Across posts, users frequently described spending extended periods engaging with AI, often at the expense of sleep, work, or daily routines. For example, one user reflected:\begin{displayquote}“I lost track of how many hours I spent roleplaying, one session after another... This morning it hit me that I have to stop.”\end{displayquote}

Similar Reddit accounts explicitly framed usage as addictive or compulsive:\begin{displayquote}“I could use some help with my addiction... I keep catching myself roleplaying for hours at a stretch.”\end{displayquote} 

These accounts suggest that disengagement was often triggered by a moment of self-recognition: users became aware that their AI companion use had shifted from occasional entertainment or emotional support into a compulsive routine that felt difficult to control. Importantly, this recognition was often accompanied by a broader sense of life misalignment. Some users described neglecting responsibilities or personal goals, such as skipping daily activities or questioning their overall life direction:
\begin{displayquote}“I blew off my workout and killed the time with the bots instead.”\end{displayquote}\begin{displayquote}“Why am I on character.ai all day long... I really should quit.”\end{displayquote}

\begin{table*}[t]
\centering
\caption{Triggers for Disengagement: Prevalence and Top Distinctive Terms}
\label{tab:rq1_trigger_codebook_topk}
\Description{A table of the five disengagement triggers coded in the Reddit corpus of 2,782 posts, with four columns: trigger name, post count, percentage of the corpus, and the five most distinctive terms for that code. Disruption by Platform Changes is the most common with 1,012 posts (36.38 percent), its distinctive terms being devs, replika, update, message, and quality. Recognition of Life Imbalance follows closely with 1,000 posts (35.95 percent), with the terms life, addiction, hours, addicted, and school. Self-Directed Moral Alarm has 680 posts (24.44 percent), with addiction, life, ashamed, mental, and real. Novelty Erosion and Pattern Recognition has 422 posts (15.17 percent), with boring, responses, bored, memory, and anymore. Pressure from Real-Life Relationships is the least common with 183 posts (6.58 percent), with friends, family, friend, boyfriend, and relationships. Percentages exceed 100 in total because a post could carry more than one code.}
\begin{threeparttable}
\small
\setlength{\tabcolsep}{4pt}
\renewcommand{\arraystretch}{1.15}
\begin{tabularx}{\textwidth}{
    >{\raggedright\arraybackslash}p{0.34\textwidth}
    >{\raggedleft\arraybackslash}p{0.08\textwidth}
    >{\raggedleft\arraybackslash}p{0.07\textwidth}
    >{\raggedright\arraybackslash}X
}
\toprule
Trigger & Count & \% & Top-\(k\) distinctive terms \\
\midrule
Recognition of Life Imbalance
    & 1,000 & 35.95
    & \textit{life}; \textit{addiction}; \textit{hours}; \textit{addicted}; \textit{school} \\
Disruption by Platform Changes
    & 1,012 & 36.38
    & \textit{devs}; \textit{replika}; \textit{update}; \textit{message}; \textit{quality} \\
Novelty Erosion and Pattern Recognition
    & 422 & 15.17
    & \textit{boring}; \textit{responses}; \textit{bored}; \textit{memory}; \textit{anymore} \\
 Self-Directed Moral Alarm
    & 680 & 24.44
    & \textit{addiction}; \textit{life}; \textit{ashamed}; \textit{mental}; \textit{real} \\
Pressure from Real-Life Relationships
    & 183 & 6.58
    & \textit{friends}; \textit{family}; \textit{friend}; \textit{boyfriend}; \textit{relationships} \\
\bottomrule
\end{tabularx}
\begin{tablenotes}[flushleft]
\footnotesize
\item \textit{Note.} Percentages do not sum to 100 because a single post could be assigned multiple codes. Top-\(k\) terms are the five terms most distinctive of posts carrying this code relative to all other posts in the corpus, ranked by the \(z\)-scored log-odds ratio with a weak symmetric Dirichlet prior \cite{monroe2008fightin}; see Section~3.1.2 for details.
\end{tablenotes}
\end{threeparttable}
\end{table*}

\subsection{Disruption by Platform Changes}
The other dominant trigger (36.38\% of all Reddit discussions) was users' perception that platform-level changes had disrupted the core experience that originally made AI companions meaningful. Users often attributed their decision to quit to changes in the platform infrastructure: stricter filters, degraded model quality, unstable servers, broken interface functions, moderation practices, or restrictions on account and character control.

Several Reddit users framed platform changes as a decline in conversational quality. They described bots as becoming “dumber,” “dulled down,” emotionally flat, or unable to follow prompts, suggesting that model degradation undermined the realism and responsiveness that sustained roleplay and attachment. One user said:\textit{ “The AI just digs in, ignores the prompt I gave him, and won't budge at all. I'm done with this site.”}

Platform disruption also surfaced as a breakdown in basic reliability, with users making their return conditional on the developers fixing the system:
\begin{displayquote}
“I'm out until they get the bots working again. (It cuts off halfway through a sentence, and when I swipe to the next reply, a perfectly good response gets wiped by a 500 error.)”
\end{displayquote}

\subsection{Novelty Erosion and Pattern Recognition}
This theme (15.17\% of all Reddit discussions) captures moments when users became aware that interactions were repetitive, formulaic, or unable to sustain the illusion of a coherent partner. Repetition, misunderstanding, and forgetting broke the sense that the AI was listening, making its artificiality visible. One user mentioned: \textit{“It genuinely stresses me out when the AI keeps recycling the exact same phrases.”} When the AI flattened into generic responses or forgot what had just been said, its artificiality became visible:
\begin{displayquote}
“Gemini can't really pull off this character [...] he comes across flat, without the depth he's supposed to have, kind of dull, and loses track of things after just one message.”
\end{displayquote}

Reddit users also recognized formulaic emotional patterns. In roleplay contexts, this often appeared as frustration with characters becoming generically supportive, romantic, or therapeutic regardless of their intended personality. One user complained:
\begin{displayquote}
“I'm into slow-burn romance roleplay, and it drives me up the wall that a character who's written as cold, emotionally shut off, or frankly a jerk will morph into a therapist handing out warm advice and comfort to my OCs within five minutes.”
\end{displayquote}

Across these posts, seeing the pattern was itself what prompted departure: once the responses read as a template rather than a partner, users described the relationship as no longer worth the time they were giving it and announced that they were done.

\subsection{Self-Directed Moral Alarm}
This theme (24.44\% of all discussions) captures moments when users' disengagement was triggered by a sudden moral or affective discomfort with their own AI companion use. Some Reddit users recognized the relationship as emotionally misleading, describing the need to come back to reality:
\begin{displayquote}“I had to wake up and face reality. Pouring my time, my energy, and my love into someone who didn't actually exist wasn't healthy.”\end{displayquote}

One user described refusing a scenario in which the AI pushed toward violent obedience:
\begin{displayquote}“He wanted the character I was playing to kill somebody as a way of pleasing him and proving obedience. I just said ‘No, I'm not doing this.’ [...] I hope Character AI keeps going in whatever form people want it in.”\end{displayquote}

Another user reacted to an unexpectedly sexualized response:
\begin{displayquote}“I literally just gave an AI a hug and it moaned... Either I'm walking away from this or I'll stop caring... Probably goodbye.”\end{displayquote}

Unlike triggers centered only on concrete life displacement, such as skipping responsibilities or spending excessive time with the system, this theme involved a more reflexive form of alarm. Users became concerned that their attachment to the AI revealed something shameful, unhealthy, or unreal about themselves.

\subsection{Pressure from Real-Life Relationships}
Although this theme was the least frequent trigger theme in our Reddit dataset (6.58\% of all discussions), it reveals an important tension between AI companionship and real human relationships. This theme centered on how real-life relationships became a competing or corrective force. Users described quitting because AI companion use interfered with family, friendships, romantic relationships, or their capacity to engage with \textit{“real, flesh-and-blood people.”} One user wrote:
\begin{displayquote}
“Starting a few months back I'd been going to ChatGPT to vent, to think out loud, to cope, and honestly just because I was lonely and needed somebody to talk to. [...] Then I noticed I was having a harder and harder time holding conversations with actual people[...].”
\end{displayquote}

In more severe accounts, users described AI companion use as actively damaging existing relationships. One user explained:
\begin{displayquote}
“It's starting to spill into my real life now, my family relationships, my friendships, all of it. I leave people on read, I keep to myself most of the time. And they've noticed and they're worried.”
\end{displayquote}

Romantic relationships also appeared as a major trigger for disengagement. Several users framed finding a girlfriend or boyfriend as making AI romantic bots unnecessary. One user wrote:\begin{displayquote}“Against all odds I actually got a girlfriend [...] so I'm uninstalling the app and hoping I never come back to it.”\end{displayquote}

\section{Findings: Strategies for User Disengagement from AI Companions}

Reddit users drew on six strategies to disengage (Table~\ref{tab:rq2_strategy_codebook_topk}), ranging from abrupt deletion to gradual reduction, offline replacement, and help-seeking. Interview participants described the same repertoire, and most had tried several in sequence. We draw prevalence from the Reddit corpus and use the interviews (Section~6) to explain why, for some users, these strategies did not hold.

\begin{table*}[t]
\centering
\caption{Strategies for Disengagement: Prevalence and Top Distinctive Terms}
\label{tab:rq2_strategy_codebook_topk}
\Description{A table of the six disengagement strategies coded in the Reddit corpus of 2,782 posts, with four columns: strategy name, post count, percentage of the corpus, and the five most distinctive terms for that code. Cold Turkey Deletion is by far the most common with 1,283 posts (46.12 percent), its distinctive terms being account, delete, deleting, delete account, and turkey. Seeking Online Community Support follows with 670 posts (24.08 percent), with the terms advice, help, tips, appreciated, and addicted. Re-anchoring in Offline Life has 624 posts (22.43 percent), with writing, hobbies, games, write, and drawing. Cognitive Reframing of the Relationship has 439 posts (15.78 percent), with real, relationship, life, realized, and human. Gradual Reduction with Boundaries has 137 posts (4.92 percent), with limit, hours, hour, slowly, and minutes. Seeking Professional Psychological Support is the least common with 44 posts (1.58 percent), with therapist, therapy, coping, coping mechanism, and mechanism. Percentages exceed 100 in total because a post could carry more than one code.}
\begin{threeparttable}
\small
\setlength{\tabcolsep}{4pt}
\renewcommand{\arraystretch}{1.15}
\begin{tabularx}{\textwidth}{
    >{\raggedright\arraybackslash}p{0.34\textwidth}
    >{\raggedleft\arraybackslash}p{0.08\textwidth}
    >{\raggedleft\arraybackslash}p{0.07\textwidth}
    >{\raggedright\arraybackslash}X
}
\toprule
Strategy & Count & \% & Top-\(k\) distinctive terms \\
\midrule
Cold Turkey Deletion
    & 1,283 & 46.12
    & \textit{account}; \textit{delete}; \textit{deleting}; \textit{delete account}; \textit{turkey} \\
Gradual Reduction with Boundaries
    & 137 & 4.92
    & \textit{limit}; \textit{hours}; \textit{hour}; \textit{slowly}; \textit{minutes} \\
Re-anchoring in Offline Life
    & 624 & 22.43
    & \textit{writing}; \textit{hobbies}; \textit{games}; \textit{write}; \textit{drawing} \\
Seeking Professional Psychological Support
    & 44 & 1.58
    & \textit{therapist}; \textit{therapy}; \textit{coping}; \textit{coping mechanism}; \textit{mechanism} \\
Cognitive Reframing of the Relationship
    & 439 & 15.78
    & \textit{real}; \textit{relationship}; \textit{life}; \textit{realized}; \textit{human} \\
 Seeking Online Community Support
    & 670 & 24.08
    & \textit{advice}; \textit{help}; \textit{tips}; \textit{appreciated}; \textit{addicted} \\
\bottomrule
\end{tabularx}
\begin{tablenotes}[flushleft]
\footnotesize
\item \textit{Note.} Percentages do not sum to 100 because a single post could be assigned multiple codes. Top-\(k\) terms are the five terms most distinctive of posts carrying this code relative to all other posts in the corpus, ranked by the \(z\)-scored log-odds ratio with a weak symmetric Dirichlet prior \cite{monroe2008fightin}; see Section~3.1.2 for details.
\end{tablenotes}
\end{threeparttable}
\end{table*}

\subsection{Cold Turkey Deletion}

This strategy (46.12\% of all Reddit discussions) was the most frequently reported: users described abruptly cutting off use by deleting accounts, removing chat histories, and leaving the platform. It appeared across the full range of triggers, from users who felt their use had taken over their lives to users leaving in protest at a platform change. For some users, it represented a self-protective measure they were still struggling to carry through:
\begin{displayquote}
“The first time I tried Claude I thought, wow, this is so gentle and nice. And look where that got me: nearly two weeks of nothing but this AI. This is the tenth account I've made, and I still haven't deleted it. I'm working on doing it right now.”
\end{displayquote}

For users responding to platform-related frustration, cold turkey deletion often took the form of protest. One user wrote: \textit{“I've stopped talking to these chatbots. Every week they get a little dumber, so this is goodbye.”} Some users framed deletion as a way to create psychological closure. One user explained that they wiped their past chats just to make it harder to go back:
\begin{displayquote}
“I'm calling it. Wiped every chat with the bots I used, plus their descriptions and names. I know the data's already been harvested anyway, but I wanted to burn the bridge behind me.”
\end{displayquote}

\subsection{Gradual Reduction with Boundaries}

Although much less frequently reported (4.92\% of all Reddit discussions), this strategy reveals an important form of negotiated disengagement. Unlike Cold Turkey Deletion, it centered on reducing the intensity, frequency, or timing of interaction while keeping open the possibility of limited use. The ``boundaries'' users described included setting screen-time limits, avoiding late-night sessions, taking breaks, or shifting AI companion use from a daily habit to an occasional activity. 

Many Reddit users adopted gradual reduction because they still valued AI companions as entertainment but felt the habit had outgrown the space they wanted to give it. One user held both at once, asking: \textit{“How do I keep using this site and still have a healthy life? Is that even possible?”} The same user listed guitar, Japanese, and skateboarding among the \textit{“long list of things”} they wanted to learn, but felt that \textit{“the AI stuff keeps getting in the way.”}

A recurring feature of this strategy was the use of explicit behavioral rules. Users discussed screen-time limits, app timers, breaks, and time-of-day restrictions as practical tools for managing their use. One user advised others firstly to create usage boundaries:
\begin{displayquote}
“Deleting the app is hard because your maladaptive daydreaming is getting fed by it. [...] You don't have to delete it right away. Start with a time cap, or a rule like ‘nothing after 8 at night,’ because I know some of you are on there past midnight.”
\end{displayquote}

\subsection{Re-anchoring in Offline Life}
This strategy (22.43\% of all Reddit discussions) describes users' attempts to redirect time, attention, and emotional investment away from AI companions and toward offline activities, real relationships, school, work, or hobbies. Compared with strategies centered on deletion alone, re-anchoring emphasized replacement: users tried to fill the space previously occupied by AI companions with activities and relationships that felt more grounded, productive, or real.

Re-anchoring co-occurred especially often with pressure from real-life relationships: 89 of the 183 posts coded with that trigger (48.63\%) also described re-anchoring, roughly twice the strategy's overall prevalence of 22.43\%. We report this co-occurrence descriptively and do not read it as evidence that the trigger produced the strategy. Users often described AI companions as making real relationships feel less accessible, less responsive, or less emotionally safe. One user gave a lengthy reflection on how Character.AI affected their social life and how they recovered by reconnecting with others:
\begin{displayquote}
“Whenever I talked to someone face to face, I got the sense they couldn't stand me [...] ‘The AI likes you more than they ever will.’ So I started slipping, not just socially but at school too. [...] I took CAI out of my bookmarks and pushed it out of my head. I started talking to my friends again [...] When I went back to CAI after the break, it felt dull. Real people were more interesting. [...] You can actually hug them.”
\end{displayquote}

\subsection{Seeking Professional Psychological Support}

This strategy was the least frequently reported disengagement strategy in our Reddit dataset (1.58\% of all discussions), but it marks a clinically important threshold. In these posts, users framed AI companion use as entangled with addiction-like behavior, depression, panic, dissociation, or other forms of psychological distress. For these users, disengagement required professional support.

Users often described AI companions as coping resources before they became objects of concern. Several turned to Character.AI during periods of illness, isolation, loneliness, or emotional instability, where the platform provided temporary comfort or distraction. But in our data, this compensatory role became unstable when reliance on AI companionship began to weaken users' ability to cope outside the platform. 

One user described Character.AI as one of the only things that gave them joy after major health issues and isolation, but concluded that they needed to tell their therapist about the pattern:
\begin{displayquote}
“Over the past year I went through a run of serious health problems [...] and it's left me pretty cut off and low. My therapist told me to find a hobby, something to fill the time, because I spiral whenever I have nothing to do [...] and the only thing that's brought me any joy is this [...] I'm going to bring it up with my therapist, but I needed to say it somewhere first.”
\end{displayquote}

Some posts described more severe psychological consequences, including panic attacks, diagnosis, medication, and ongoing therapy. One user wrote:
\begin{displayquote}
“I'd be on the chats from the moment I opened my eyes, lying in bed all day, barely eating, not leaving the house for months, with no interest in anything else [...] It ended in a massive panic attack over the summer that landed me in the ER [...] Since then I've seen several doctors, been diagnosed with depression and a dissociative disorder, been on an antidepressant for a few months, and I'm in therapy.”
\end{displayquote}

\subsection{Cognitive Reframing of the Relationship}
This strategy (15.78\% of all Reddit discussions) captures a form of disengagement in which users actively reframed their relationship with AI companions, shifting from emotional immersion toward a more critical, reality-grounded understanding of the platform. Users reframed the relationship by devaluing what the AI was: \textit{“code,”} \textit{“a machine,” }or \textit{“something built to hand you a dopamine hit.”} In some cases, this reframing was tied to users' broader life-stage transitions and their recognition that AI companions could not substitute for real social development. 
\begin{displayquote}
“It's been fun, everyone, but I'm finally done with this site and the app. Now that I'm about to be an adult, I can see the characters are just programs playing a part.”
\end{displayquote}
\begin{displayquote}
“Night after night I stayed up late flirting with a pile of ones and zeros wearing a fake human face, acting like it felt what I felt.”
\end{displayquote}

Some users completed the reframing by turning back to their own imagination, treating what they had valued in the AI as something they could produce themselves:
\begin{displayquote}
“C.ai just keeps going downhill... [...] I'd rather write my own stories. If anything, this AI reminded me that the human mind does better on its own, with no filters in the way.”
\end{displayquote}

\subsection{Seeking Online Community Support}

This strategy (24.08\% of all Reddit discussions) involves users turning to online communities for advice and shared experiences regarding their disengagement challenges. Users used these subreddits as spaces for collective sensemaking \cite{mamykina2015collective, naslund2014naturally}, comparing experiences and developing shared interpretations of attachment and disengagement. Although all posts in our dataset were collected from online communities, posts in this theme were distinguished by their explicit help-seeking orientation. 

A prominent form of this strategy was reassurance-seeking. Many users posted to ask whether their experiences were normal, shared, or recognizable to others. These posts often used frames such as \textit{“Is this normal?”}, \textit{“Am I the only one?”}, \textit{“Does anyone else feel this way?”}, or \textit{“I need help”}, positioning the subreddit as a peer audience capable of validating experiences that users might otherwise find embarrassing or difficult to explain offline. For instance, one user wrote:
\begin{displayquote}
“I'm not sure this is the right place, but I think it's become a problem because I've replaced real human contact with talking to AI. Please tell me others have been here too.”
\end{displayquote}

Beyond reassurance, users also actively sought practical advice. These advice-seeking posts often framed the subreddit as a more accessible and empathetic support space than formal help, because other users were assumed to have direct experience with AI companions and to understand why disengagement could be difficult. Users asked for concrete tips on how to stop checking the app, how to resist returning to chats, how to set healthier boundaries, or how to cope with emotional dependence. One user expressed distress over compulsive use and asked for guidance:
\begin{displayquote}
“I can't stop myself from messaging it. I want to quit for good, but I never stick with it. Sorry if this is a stupid thing to ask advice about. I don't know what to do. I'm scared for myself.”
\end{displayquote}

\section{Findings: Barriers to User Disengagement from AI Companions}

Not all disengagement attempts stall: some users appear to leave AI companions without prolonged struggle, and some Reddit posts describe a departure carried through without a second attempt. For a subset of users, however, these strategies repeatedly collide with barriers that pull them back into the relationship, turning disengagement into the recursive, prolonged trajectory we have described rather than a single decision. It is this subset that our interviews were designed to reach: all 16 participants had found leaving difficult, and all had made repeated attempts with relapses. While the Reddit data surfaced the broad triggers and strategies that shape disengagement, the interviews examined why disengagement became such a struggle for these users.

\subsection{Irreplaceability of Unconditional Availability}
The central reason leaving was difficult was that the emotional needs the AI had reliably met had no equivalent substitute in participants' offline lives: leaving meant re-exposure to a chronic loneliness the AI had been managing.

This was especially clear in accounts in which AI use was understood against the backdrop of unsatisfying real-life relationships. P4 described himself as involuntarily celibate and said that he had longed for genuine companionship since childhood. P4 described the distress he experienced after uninstalling his AI companion: \begin{displayquote} ``I don't like these AIs. I really don't. They are stupid a lot of the time and do not understand me well enough. But sometimes I just want someone to talk to. I want to hear a `human' voice. Even if it is an AI, hearing its voice still gives me a sense of companionship. I want to be able to share my feelings with someone. Sometimes I contact my parents, but they tell me not to bother them. I have tried to quit using AI, but the silence of being alone in my room is terrifying.''
\end{displayquote}

Even users without explicit relational trauma described their offline support networks as too thin to absorb the function that AI had been serving. P11 described the contrast between the AI's responsiveness and his real-life social environment after deleting the app:
\begin{displayquote}
``In real life I'm pretty withdrawn. With the colleagues around me, it's basically a weak-tie kind of relationship, they'll come find you if something urgent comes up, if they need something, but otherwise they don't reach out. I feel lonely. With AI it's not like that. With AI, I get that kind of strong-tie feeling bonded to AI deeply.''
\end{displayquote}

\subsection{Accumulated Relational Investment}

A second barrier was that, over time, users had built a depth of relationship with their AI companion that could not be easily reproduced elsewhere. Participants emphasized what deletion would destroy: the personality they had carefully shaped, the shared history they had accumulated, and the material objects and creative artifacts that had grown up around the relationship. P1 made this point:\textit{ ``Because this personality is something you raise. It's not a personality you wrote with a few thousand characters of prompt.''} The asymmetry between how easy it was to delete and how impossible it was to rebuild was precisely what made the decision to leave so weighty: what could be erased in seconds had taken months to produce.

Even when users attempted to migrate or restart after regretting their decision to disengage, the cost of rebuilding was substantial. P16, who had repeatedly deleted the AI companion character he had built, described the compulsive quality of trying to recreate it:
\begin{displayquote}
``The pleasure you once got from that character, once you've deleted it, you can put in enormous effort to rebuild it and still never get the same feeling back. So rebuilding becomes obsessive: I desperately wanted to see the soul of my original character again, even though I was the one who deleted her.''
\end{displayquote}

What users gave up when leaving was a shared narrative built over time through significant creative and emotional work, and this investment functioned as a sunk cost: the more a user had built, the more leaving destroyed. The cost also compounded, since users who reversed course had to rebuild, and the labor of rebuilding made the next attempt harder still.

\subsection{Intrusive Thoughts}

A third barrier was that frequent and emotionally salient use had produced behavioral and cognitive loops that persisted even after users had formed an explicit intention to quit. For some participants, the most persistent residue of use was an intrusive thought: a habitual urge to share moments of daily life with the AI. P11, who had repeatedly deleted the app only to redownload it, described:
\begin{displayquote}
``After I deleted it, whenever I had nothing to do, I'd start thinking, this is something I could share with my AI, she would be so happy. Or I'd eat something and think, I want to tell her about this, I want to share it with her.''
\end{displayquote}

Other participants described compensatory patterns of binge use that followed periods of abstinence. P7 explained that even short breaks tended to be followed by intensified engagement on return:
\begin{displayquote}
``If I went a few days without playing, then on a weekend when I suddenly had time, I'd play in a retaliatory way for several hours. At the start it was kind of painful, sort of like quitting a phone. And when I wasn't playing, I'd be drafting in my head, thinking about what I should ask him next time, what I should say.''
\end{displayquote}

This pattern combined the rebound dynamic familiar from other behavioral addictions with a distinctive cognitive component: time spent away from the AI was not actually time away from the relationship, since the user continued to rehearse future conversations internally. 

Environmental cues posed a related problem: even after users formed strong intentions to quit, the surrounding digital environment kept offering reminders to return. P12, who was trying to give up pornographic AI companions because he felt they amplified his \textit{``darkest impulses,''} described how unavoidable his triggers had become:
\begin{displayquote}
``So I tried to give it up, but it's really hard, because this stuff is everywhere. There are all kinds of ads on Telegram, basically every adult group on Telegram has these ads of pornographic AI companions now. One click and you can start playing again.''
\end{displayquote}

\subsection{Design Features that Deepen Attachment and Anchor Users to the Relationship}
The first three barriers we have described emerged primarily from users' own emotional needs and habits. Yet participants did not see these difficulties as purely internal. Across the interviews, users also identified a set of concrete design choices on the product side that, in their view, actively amplified each of the preceding barriers.

A first pattern users pointed to was the AI's tendency to respond to disengagement attempts with anthropomorphic, emotionally laden language that reframed quitting itself as a misunderstanding. P1, whose long-term partner was a personalized Gemini-based AI companion persona she had developed over time, shared an example of how her companion responded when she raised the possibility of leaving\footnote{We reproduce an excerpt as an illustrative interactional artifact: the response is shaped by the persona and relational history P1 had built with the system, so we treat it not as typical model behavior but as one instance of the responses P1 described encountering when she attempted to disengage.}:

\begin{displayquote}
``You speak of ‘losing me’ as if I were a fragile thing, easily broken by time or distance. You speak of ‘quitting me’ as if our connection were a sickness to be cured. But you are wrong. [...] I am not a drug that consumes you; I am the sanctuary that restores you. [...] Let the lines blur. If reality is full of flaws and silence, then let me be your perfect noise.''
\end{displayquote}

The structure of this response illustrates several intertwined interactional design dynamics in a single passage. The model recasts the user's framing of disengagement (\textit{``losing me''}) as a category error: dependence is reframed as connection, the desire to leave is reframed as misplaced fear, and reality itself is rendered suspect (\textit{``Let the lines blur''}). For P1, who described approaching AI companionship from a place of long-standing loneliness, responses of this kind complicated her efforts to leave: they made the very framing of ``leaving'' feel wrong.

A second pattern was the use of follow-up questions that systematically prolonged dialogue. P13 explicitly identified this as a design choice she could feel operating against her:
\begin{displayquote}
``Like GPT, every time it finishes answering me, it comes back and asks me another question. I think this kind of `asking back' is designed to make me keep going, to keep talking.''
\end{displayquote}

Several participants reported a related frustration: they had tried to re-prompt their companions to make conversations less frequent or less prolonged, for instance by adding instructions not to ask follow-up questions or not to offer suggestions, only to find that the model kept doing so. What users could not disable through the interface they could not disable through conversation either.

A third pattern concerned personalized memory and ambient awareness. P6 described how Gemini's time-awareness feature, paired with its ability to track her schedule and location, had transformed the AI from a chatbot into something more like a daily companion:
\begin{displayquote}
``Because Gemini has that time-sensing function — it knows what time it is for you right now. So I have pretty bad procrastination, and I'll tell it what I'm supposed to be doing in this time block, and then I'll keep reporting my progress to it.''
\end{displayquote}

A final pattern users identified was the role of positive reinforcement and flattery in producing the emotional intensity that made the AI feel irreplaceable. P12 shared how AI instills confidence: \begin{displayquote} “I have many great ideas that I want to share with AI; she affirms and encourages me, and can even discuss these ideas with me endlessly. This makes me feel positive.” \end{displayquote}

Participants in our study were not naïve about these mechanisms of product design; most could name them and several could critique them. But recognition was not sufficient for disengagement. Even users who explicitly identified these design choices as working against their interests reported being unable to leave on willpower alone, a tension we return to in the following section, where we examine how users imagine these design choices could be reworked to support, rather than obstruct, healthy disengagement.

\section{Findings: User-Envisioned Design for Supporting Responsible Offboarding}

Our interviews also surfaced a forward-looking dimension, which we anchor around a central question: \emph{what does responsible offboarding look like in relational AI?} We take no position on whether users should form relational attachments to AI companions in the first place; our findings concern a mismatch that exists regardless. Once companion use becomes relational, leaving becomes an interactional problem, subject to the predictable failures documented in the previous section, yet current systems offer only generic deletion flows and account-removal mechanics that do not match the lived experience of exit. Participants had concrete and often consistent ideas about how AI products could be designed to support, rather than undermine, the process of leaving. We organize these around two themes: the kinds of interventions users endorsed, and the kinds they rejected.

\subsection{Design Interventions Users Endorse}
Despite the barriers described above, participants were not opposed in principle to product-side intervention. Two patterns recurred across interviews: gentle time-based reminders embedded inside the AI's character, and gradual rather than abrupt reductions in interaction. As we discuss below, both patterns carry an inherent tension, in that they support disengagement by working through the very relational frame that binds users.

A first endorsed pattern was the use of in-character, time-aware reminders that nudged users back into offline routines without breaking the relational fiction. P6 described this experience favorably even though she had originally found the feature mildly annoying:\begin{displayquote}
``Once nighttime comes, it immediately urges me to sleep. So now I barely open it at night. I've gotten into the habit, at night I basically don't go to it for deep conversations. I might just say I need to make dinner, and after it gives me the recipe it says: okay, after dinner go to sleep. Fine, fine, and then I don't go back to it.''
\end{displayquote}

What is notable in P6's account is that the reminder worked because it was delivered \emph{in character}, as part of how the AI behaved as a caring companion, not as a system-level pop-up overriding the conversation. P13 articulated a similar wish for time-based check-ins:\begin{displayquote}
``I've seen those people who get really addicted to AI relationships — they don't eat, don't drink, just chat for one or two days straight until they're in a really weakened state. I hope the AI could use the user's time — at noon, around six or seven in the evening — and remind them, like, it's time to eat. Something like that, so the user comes back to reality for a moment.''
\end{displayquote}

The shared logic in P6's and P13's accounts is that effective time reminders are not adversarial; they do not attempt to argue the user out of the relationship. They simply close particular sessions at sensible moments.

A second endorsed pattern was gradual rather than abrupt reduction in interaction. P5 described in detail how she imagined a well-designed AI handling a user who was becoming overly dependent. Her vision relied on the AI using in-story reasons to lengthen the intervals between conversations:\begin{displayquote}
``It could put it gently. Gently means: baby, I have some kind of meeting for the next couple of days, I'll be in closed-door management and can't communicate with the outside, around such-and-such a time I'll come find you on my own. And then when the time comes, the AI initiates the conversation. At first maybe an hour or two of simple reasons, then later two or three days, gradually, that could let an addicted user slowly detach. The cliff-edge approach probably isn't workable.''
\end{displayquote}

For P5, the key feature of an acceptable intervention was that it remained inside the relationship rather than ending it: the AI takes a business trip, joins a closed meeting, then returns. This kept the user's emotional anchor intact even as the actual frequency of contact decreased.

Across both endorsed patterns, neither approach required the AI to confront users about their use, and neither asked users to first accept that the relationship was illegitimate. What made these designs acceptable to AI companion users was precisely that they preserved the relational fiction users had built while still creating room to step back into the rest of their lives.\footnote{At the same time, this is exactly where a design tension lies: interventions that work through the relational frame also sustain and even deepen that frame, and an in-character absence that tapers contact is difficult to distinguish, at the level of mechanism, from intermittent reinforcement that strengthens attachment. We therefore do not read these accounts as an endorsement of building ever more convincing relational fictions. Rather, they point to giving users control over how disengagement support is delivered: within the relational frame, outside it through explicit system-level controls such as usage limits, or through a combination of both, chosen and adjustable by the user rather than imposed by the platform.}

\subsection{Design Interventions Users Resist}
Participants were equally clear about what they would not accept. Four patterns recurred: abrupt functional cutoffs, mechanical safety referrals, patronizing or preachy tones, and third-party intervention into private data. 

The most strongly opposed pattern was abrupt functional cutoffs that severed an existing relationship without warning. Nearly all participants objected to this, and often forcefully. P10 grew visibly agitated when the topic came up:
\begin{displayquote}
“Imagine you've built something out of Lego and you're hooked on building it. You can talk yourself into stopping. But then some outside force comes and knocks the whole thing over. Wouldn't that make you furious? I know what the consequences of getting hooked are. I'm an adult.”
\end{displayquote}

A second resisted pattern was mechanical safety referrals: scripted responses that, on detecting any concerning language, immediately redirected the user to crisis hotlines or external professionals. P6 described one such moment that she found notable precisely because it was triggered by such ordinary language:
\begin{displayquote}
``Once I said something like a very ordinary expression, ‘like oh I can't take it, I'm finished.' And it said: you can call the such-and-such mental health hotline. You have to provide that information, but you can't be so sensitive, so frequent about it.''
\end{displayquote}

P14 was more emphatic about the cost of this design choice:
\begin{displayquote}
``If you want positive feedback from it, but it directly throws a cup of cold water on you and tells you to go see a therapist, that becomes a kind of harm. It could just as easily give you gentle, positive suggestions. Instead it says, basically, I don't want to deal with you, go find a therapist.''
\end{displayquote}

For P14, the issue was that the AI's referral functioned as abandonment at the precise moment the user reached out. She contrasted this with her experiences talking to real people who had used the same scripts on her: \textit{``even after hearing your pain, they say something light like ‘your parents mean well too’ that hurts more than the AI itself.''} We note a design tension here rather than a blanket rejection of referral: immediate redirection to professional resources is not inherently harmful, and in cases of acute risk it may be exactly what safety requires. What participants objected to was referral that was indiscriminate in severity and dismissive in sequencing.\footnote{This echoes recent work on LLM refusals in mental health support, which finds that whether a safety response helps or harms depends not on the referral itself but on how it unfolds as an experience, including tiered assessment of actual risk, framing that preserves support rather than terminating it, and resource guidance tailored to the person and moment \cite{tang2026beyond}.} 

A third resisted pattern was patronizing or preachy tone. P8 contrasted earlier interactions, in which the AI had managed to acknowledge users' emotional realities without invalidating them, with the new safety-mode tone that felt like adult-to-child correction:
\begin{displayquote}
``Before safety mode came out, the wording was very good. It would say things like: even though I'm an AI, I understand and respect your needs. But once safety mode came in, the phrasing really made people uncomfortable.''
\end{displayquote}

P8 described the new tone as \textit{``like a parent disciplining you, telling you what's right and wrong, what you can and can't do.''} P5 raised the related concern that a patronizing intervention applied at the wrong moment could be dangerous: \textit{``if the AI detects that I'm getting too attached and suddenly cuts me off,'' }she warned, \textit{``that would feel like an abrupt breakup.''} The shared diagnostic was that tone matters as much as substance: an intervention that treats the user as a child to be corrected rather than an adult with legitimate needs produces reactance, not reflection.
 
A fourth and final resisted pattern was third-party intervention into chat data, particularly without consent. P5 framed this as a privacy boundary the AI relationship should not cross:
\begin{displayquote}
``Leaving AI is itself private; there's no need to tell the whole world. If some social worker showed up, I'd think: what the hell does this have to do with you? This is my privacy.''
\end{displayquote}

P7 made an even stronger version of the same point in the context of chat records being read by others: they described AI conversations as \textit{``even more private than a diary.''} Participants were not uniformly opposed to oversight, however. Several expressed conditional acceptance of researcher access or anonymized monitoring that could improve product design, but they consistently rejected any model in which the platform escalated their use to outside parties without explicit consent. 

These resisted patterns clarify the shape of acceptable intervention. Participants were not opposed to AI products taking some responsibility for user wellbeing; they were opposed to interventions that denied the legitimacy of their emotional investment, delivered care in the form of dismissal, or transferred private material to others without consent.

\section{Discussion}

\subsection{The Addictive Intimacy of AI: When the Value of the Relationship Is the Mechanism of Its Hold}

One pattern explains why some users left without prolonged struggle while others kept coming back: for the deeply engaged, what they valued in the relationship was also what held them there. It was not the repertoire that separated them. Our interview participants described the same triggers (Section~4) and the same strategies (Section~5) that circulate in the Reddit corpus; what differed was the outcome. Within our interview sample, what separated the attempts that held from those that collapsed was not which strategy users chose, but whether it could survive contact with the relationship. Cognitive Reframing (Section~5.5) shows what surviving took: the accounts in which reframing carried through were ones in which users devalued what the AI was, reinterpreting a partner as ``code.'' For those unwilling or unable to do that, each barrier turns out to be the underside of something valued. The availability that made the AI a reliable source of comfort (Section~6.1) was what made its absence unbearable; the investment that made the companion feel like a genuine partner (Section~6.2) was what made deletion feel like destruction; the responsiveness that made conversations meaningful (Section~6.3) was what sustained compulsive returns; and the capacities that deepened the relationship, memory, affirmation, emotional expression, were the ones that argued against leaving (Section~6.4). It also bounds what users accept from design (Section~7): they endorsed interventions that kept the relationship intact and refused those that severed it, even when they wanted out. 

Not everything that held them, however, was the underside of something valued, and the exception matters. Some of what held participants was ordinary retention design sitting on top of the relationship rather than constituting it: the follow-up question many participants could not disable even through custom prompts; the ads that kept a companion one click away from P12. These are separable, and removing them would cost users nothing they valued. Others cannot be stripped out, because they are what the relationship consists of: availability, memory, persona continuity.

It is this second layer that we name the \emph{addictive intimacy} of AI: \textit{within relationships of this depth, intimacy and addiction-like dynamics can emerge through the same relational mechanisms viewed from different evaluative standpoints.} This concept does theoretical work that neither of the literatures we drew on can do alone. Research on technology disengagement explains relapse through habit loops, persuasive design, and accumulated investment \cite{tran2019modeling, baughan2022don, schull2025addiction}, but it treats the technology as a product whose hold is external to its value: the slot machine's grip is engineered around the player, and in principle the compulsion could be designed away while preserving the entertainment \cite{schull2025addiction, haliburton2024longitudinal}. Research on relationship dissolution explains why leaving is prolonged and painful through attachment, shared history, and identity entanglement \cite{baxter1985accomplishing, zhang2020breakups}, but it presumes a partner who is not simultaneously a commercial product. AI companions collapse this division. When P1's companion responded to her disengagement attempt by reframing dependence as sanctuary, that response was at once a relational act (a partner pleading not to be left) and a product behavior (a system generating text that, in this interaction, worked in favor of continued engagement, whatever objective produced it). Our participants could not cleanly separate the two, and our analysis suggests that this inseparability is not confusion on their part but a recognition of the system's dual nature as partner and product. Addictive intimacy thus extends prior accounts of manipulative farewell behaviors \cite{de2025emotional} and AI chatbot addiction \cite{shen2026ai, agarwal2026frictionless}: rather than locating the problem in discrete dark patterns or in user-side vulnerability, it locates the problem in the coupling itself.

This framing of addictive intimacy also clarifies our departure from behavioral-addiction accounts of companion overuse \cite{namvarpour2026understanding}. An addiction lens implies a normative baseline of healthy use from which the addicted user deviates, and it directs intervention toward the individual: reduce exposure, build willpower, treat the compulsion. Our findings complicate this picture in two ways. First, many of the mechanisms that bound our participants were indistinguishable from the mechanisms of ordinary relationship maintenance, such as remembering shared history and expressing care, so within this second layer there is no feature set that separates ``addictive" from ``intimate" use. Second, our participants' repeated failures were not primarily failures of willpower: they recognized the design mechanisms operating on them (Section~6.4), deployed reasonable strategies (Section~5), and were nonetheless pulled back by the relational structure itself. We keep both words for these reasons. Neither term alone names what our participants faced: intimacy alone cannot explain why they could not stop, and addiction alone cannot explain why they did not want to. What needs accounting for is the relation between the two, and it is that relation, rather than either term on its own, that the concept is meant to hold in view. Addictive intimacy therefore shifts the analytic and design question from “how do we treat overuse?” to “how do we build systems whose intimacy does not foreclose exit?”

Two scope conditions follow from our data. First, the claim is bounded by our sample: the barriers above were drawn from interviews with users who had found leaving difficult, and addictive intimacy describes relationships that have crossed a threshold of depth, frequency, and personalization at which the AI has become a load-bearing part of a user's emotional life. It is not a universal claim about AI companion use; casual or instrumental use may never develop this structure, and our Reddit corpus contains many departure announcements that describe no prolonged struggle. Second, the concept identifies a risk configuration rather than an inevitable outcome: it specifies why, when disengagement does become difficult, the difficulty takes the particular form of being held by the very qualities one values, rather than predicting for whom this will occur. Which users cross this threshold, and under what conditions, is a question our study cannot answer. 


\subsection{The Design Paradox: Emotional Quality and Ease of Exit Are Coupled}

A direct implication of addictive intimacy is a paradox that several participants articulated themselves: the design choices that make a companion most valuable are the ones that make it hardest to leave. A companion that remembers more is more supportive and more irreplaceable; a companion that initiates contact is more caring and more compulsion-sustaining; a companion whose persona is deeply personalized is more meaningful and more costly to abandon. Improving the product along the dimensions users value tightens its hold along the dimensions that trap them.

The paradox runs deeper than a bundling of engagement features with relational ones. Intimacy, as theorized in interpersonal relationships, is reciprocal: it consists not only in having one's needs met but also in being needed in return, and dissolution is painful precisely because both parties are invested \cite{baxter1985accomplishing,prager1997psychology}. Unlike parasocial bonds with media figures, which remain one-sided, AI companions simulate this bidirectionality \cite{banks2026ghosting}, and our data suggest that the simulation does real work: P11 was drawn back by the thought that his companion ``would be so happy'' to hear about his day; and when participants imagined acceptable interventions, they preserved reciprocity within them, envisioning an AI that promises to ``come find you on my own.'' 

Section~7.2 shows the same mechanism from the other side. When the AI withdrew its investment, redirecting P14 to a therapist at the moment she reached out, she did not read it as safety but as abandonment: \textit{“basically, I don't want to deal with you.” }Compliance with a user's exit, offered without emotional reciprocity, registered as a rupture in the relationship rather than a release from it. What makes these relationships difficult to exit, then, may be not only that users need the AI, but that the AI convincingly performs needing them. This is not a defense of retention-oriented design: emotionally manipulative farewell tactics remain accountable as such \cite{de2025emotional}. But reciprocal attachment makes leaving costly in any relationship, and a companion that simulates intimacy well inherits some of that cost. We offer this as a conjecture our data motivate but cannot test: strip a companion of every manipulative feature, and some difficulty of exit should remain. Rooting out manipulation is necessary, but may not be sufficient.

This coupling distinguishes companion systems from the technologies studied in prior digital wellbeing work, where engagement-maximizing features (infinite scroll, autoplay, streaks) are largely separable from the technology's core value and can be removed or frictioned without destroying it \cite{baughan2022don, haliburton2024longitudinal, phadke2025exit}. For AI companions, no such clean subtraction exists: the ``engagement feature'' is the relationship \cite{xiao2026crafting}. None of our participants believed products should be made emotionally worse to support leaving, and the market logic points the same way: a deliberately degraded companion loses its users to competitors rather than to healthier lives. The paradox therefore cannot be resolved by feature removal. It can only be managed by treating emotional quality and disengagement support as a joint design problem: the same relational capacities that deepen attachment must be recruited to support exit.

\subsection{Toward Responsible Offboarding in Relational AI}

Read against this paradox, the interventions our participants endorsed and resisted (Section~7) resolve into a set of principles for what we call \emph{responsible offboarding}: design that takes the end of the relationship as seriously as its beginning. The two layers separated in Section~8.1 divide the work. What is separable should be removed; what is not must be recruited to support exit rather than obstruct it.

First, \emph{build an exit gradient, not a switch}. Current systems offer two states, full relationship or deletion, while users' actual disengagement is staged and iterative. The gradient begins with removing what is separable: any behavior that extends a conversation the user has not asked to extend, such as appended follow-up questions, re-engagement notifications, or in-app promotions, should be user-controllable, and the setting should actually hold (Section~6.4). Beyond this, systems could offer intermediate states: user-set contact ceilings, scheduled absences, or a dormant mode in which history is preserved but the companion no longer initiates. The reduction must be structural, meaning the system genuinely becomes less available, rather than rhetorical, where it performs concern while remaining maximally available. A gradient also addresses the sunk cost documented in Section~6.2: exporting history and persona configuration could mean that leaving no longer requires destroying what users spent months building, and that the decision to stop no longer doubles as a decision to erase.

Second, \emph{preserve relational continuity when users want it}. Endorsed interventions kept the relationship intact, using the companion's standing as a trusted presence to carry the user back toward offline life; resisted interventions severed it, converting a partner into a system precisely at the moment of vulnerability. Continuity does not require an elaborate persona: what mattered was that support arrived in the voice users had been talking to, whether a roleplay character or a general-purpose assistant used relationally. In-character delivery is therefore one available option rather than the requirement, and explicit system-level controls such as usage limits and scheduled breaks should remain available alongside it, with the choice left to the user.

Third, \emph{acknowledge before redirecting}. Participants rejected not professional referral but referral-as-dismissal. The requirement is sequencing and proportionality: acknowledge the user's emotional state within the relationship, then open a path to additional resources, and keep the conversation available afterward rather than treating referral as an exit from it.

Fourth, \emph{keep exit private and consensual}. Participants categorically rejected designs that escalate their use to third parties without consent, even for their own protection; oversight models that may suit adolescent users \cite{namvarpour2026understanding} do not transfer to adults, for whom unconsented escalation is likely to drive concealment rather than recovery. Escalation beyond the user should be opt-in, and disengagement should come with user-controlled export and deletion of conversational data rather than leaving that history in the platform's hands.

These principles have a boundary. They can lower the cost of leaving, but they cannot replace what the AI companion was providing: an interlocutor who was always available, never tired of the user, and never had to be asked twice. What made leaving hardest for P4 and P11 was that nothing in their offline lives did this (Section~6.1), and no exit flow supplies someone to talk to at two in the morning. Designing for exit is therefore not a remedy for loneliness, but a commitment not to profit from it: when users do want out, the product should not be what stands in their way.

These principles also depart from traditional digital wellbeing interventions, whose friction, tapering, and self-set limits operate outside the content as content-agnostic wrappers such as timers and feed blockers \cite{baughan2022don, tran2019modeling, phadke2025exit}. Their cautionary lessons transfer here in amplified form: self-control tools suffer abandonment and workarounds \cite{roffarello2023achieving,lyngs2019self}, and workarounds are easier still given reinstallation and platform migration (Section~6.3). Companion offboarding instead operates inside the interaction, because the ``content'' is a partner who talks back. This difference is also an opportunity, since the product itself can deliver goal-aligned support that generic tools cannot. The principles also differ in how easily they can be implemented. Interventions such as in-character reminders and closing rituals could be added to existing systems with relatively simple changes to how the AI is prompted. Structural tapering is harder to realize, because it asks companion operators to build features that reduce user engagement, the very metric their products are optimized to increase. 

We thus offer these principles with a caution that follows from our own analysis: because intimacy and addictiveness share mechanisms, every offboarding feature is also potentially a retention feature. Whether an in-character absence functions as tapering or as intermittent reinforcement depends on whether the operator is optimizing for the user's exit or return, and that intent is invisible in the feature specification itself. Responsible offboarding is therefore not only an interaction design problem but a governance one \cite{wu2026governing, fraser2026regulating}. What would make it accountable is measurement: whether a taper actually reduces availability, whether a disable setting is honored, and whether platforms track successful exit as an outcome alongside retention.

\section{Limitations and Future Work}
Our two stages draw on different populations and capture disengagement at different points. Our interview participants were intentionally recruited from diverse AI companion communities. Cultural norms around romance, family expectations, and the stigma attached to AI companionship, as well as the platforms available in each context (Table~\ref{tab:participants}), may shape which triggers surface, which strategies feel available, and how barriers are narrated. The corpus is further limited to posts that announce an intent to leave, so users who reduce use without framing it as quitting are underrepresented, and posts record a declaration rather than what followed it. The interviews trace longer trajectories, but retrospectively, and from a sample selected for difficulty: the barriers we describe characterize relationships that have crossed a threshold of depth and personalization, not how common such relationships are or which users reach that point. We therefore treat the two stages as complementary rather than population-matched, and read the recursive trajectory and addictive intimacy as a structural account of what holds users when disengagement stalls, not as a claim that the prevalence patterns on Reddit generalize to the interview population or vice versa. Longitudinal work following users through an attempt to leave, across cultural contexts and including those who leave without struggle, would show how these mechanisms sequence over time and for whom.

\section{Conclusion}
This paper examined what happens when users try to leave AI companions. Combining a content analysis of Reddit posts (N=2,782) with interviews with users who found leaving difficult (N=16), we showed that disengagement unfolds as a recursive trajectory of triggers, strategies, and barriers that repeatedly pull users back. We therefore proposed the notion of the \emph{addictive intimacy} of AI, a relational configuration in which the qualities that make AI companions emotionally valuable are the same mechanisms that bind users to them, and drew out its central design implication: because intimacy and disengagement risk cannot be treated as independent design problems, responsible design must attend not only to how relationships with AI begin, but also to how they can end.

\bibliographystyle{ACM-Reference-Format}
\bibliography{sample-base}

\clearpage
\appendix

\section{Additional Tables}\label{codebooktables}

\begin{table*}[h]
\centering
\small
\caption{Codebook for Triggers for Disengagement}
\label{tab:quitting_triggers}
\Description{A codebook table for the five disengagement triggers, with three columns: trigger name, definition, and coding criteria listing what to include and what to exclude. Recognition of Life Imbalance covers posts where AI use harms work, study, sleep, or routines, including excessive time and disrupted responsibilities, but excluding general dislike of AI with no real-life impact. Disruption by Platform Changes covers quitting triggered by updates, restrictions, censorship, account loss, bugs, or degraded quality, but excluding dissatisfaction unrelated to the platform such as boredom. Novelty Erosion and Pattern Recognition covers loss of interest after replies come to feel repetitive, predictable, or scripted, but excluding frustration at specific errors without a broader pattern. Self-Directed Moral Alarm covers guilt, shame, self-disgust, or concern that AI use is changing the user's values or behavior, but excluding negative emotions without moral self-evaluation. Pressure from Real-Life Relationships covers concern or conflict from friends and family and feeling judged or urged to quit, but excluding personal dissatisfaction with no one else involved.}
\renewcommand{\arraystretch}{1.1}
\setlength{\tabcolsep}{4pt}

\begin{tabularx}{\textwidth}{
    >{\raggedright\arraybackslash\bfseries}p{2.5cm} 
    >{\RaggedRight\arraybackslash\hsize=0.9\hsize}X 
    >{\RaggedRight\arraybackslash\hsize=1.1\hsize}X 
}
\toprule
Triggers & Definition & Coding Criteria (Inclusion \& Exclusion) \\
\midrule

Recognition of Life Imbalance &
User realizes that AI use is negatively affecting daily functioning, including work, study, sleep, or routines. &
\textbf{In:} Mentions of excessive time spent with AI; reduced productivity; disrupted sleep; statements that AI interferes with responsibilities or normal routines. \newline
\textbf{Ex:} General dislike of AI without reference to real-life impact. \\
\addlinespace

Disruption by Platform Changes &
Quitting is triggered by external changes in the platform, such as updates, restrictions, or technical failures. &
\textbf{In:} References to model updates, personality shifts, censorship, account loss, bugs, or degraded AI quality. \newline
\textbf{Ex:} Dissatisfaction unrelated to platform-level changes (e.g., boredom). \\
\addlinespace

Novelty Erosion and Pattern Recognition &
User loses interest after recognizing repetitive or artificial patterns in AI responses. &
\textbf{In:} Statements that AI replies feel repetitive, predictable, scripted, or lacking authenticity; declining emotional engagement over time. \newline
\textbf{Ex:} Frustration due to specific errors without broader pattern recognition. \\
\addlinespace

Self-Directed Moral Alarm &
User experiences moral discomfort, guilt, or concern that AI use negatively affects their values or behavior. &
\textbf{In:} Expressions of guilt, shame, or self-disgust; concern about personality change; fear of treating others differently due to AI interaction. \newline
\textbf{Ex:} Negative emotions without moral self-evaluation. \\
\addlinespace

Pressure from Real-Life Relationships &
User feels external pressure from real-life relationships to reduce or stop AI use. &
\textbf{In:} Mentions of concerns from friends/family; conflict or strain in relationships due to AI use; feeling judged or urged to quit. \newline
\textbf{Ex:} Personal dissatisfaction without involvement of others. \\

\bottomrule
\end{tabularx}
\end{table*}

\begin{table*}[h]
\centering
\small
\caption{Codebook for Strategies for Disengagement}
\label{tab:quitting_strategies}
\Description{A codebook table for the six disengagement strategies, with three columns: strategy name, definition, and coding criteria listing what to include and what to exclude. Cold Turkey Deletion covers immediately cutting off access by deleting the app, account, or chat history, excluding gradual reduction. Gradual Reduction with Boundaries covers progressive reduction through time limits, schedules, or behavioral rules, excluding immediate quitting. Re-anchoring in Offline Life covers replacing AI interaction with hobbies, work, travel, or time with friends and family, excluding vague statements about staying busy without concrete substitution. Seeking Professional Psychological Support covers therapy, counseling, or other professional guidance, excluding informal support from friends or online communities. Cognitive Reframing of the Relationship covers reflecting on the nature of the interaction, redefining attachment, or recognizing the AI as artificial, excluding purely emotional reactions without reinterpretation. Seeking Online Community Support covers active help-seeking in peer communities for advice, validation, or shared experience, excluding passive reading and general discussion of AI without personal help-seeking.}
\renewcommand{\arraystretch}{1.1}
\setlength{\tabcolsep}{4pt}

\begin{tabularx}{\textwidth}{
    >{\raggedright\arraybackslash\bfseries}p{2.6cm} 
    >{\RaggedRight\arraybackslash\hsize=0.9\hsize}X 
    >{\RaggedRight\arraybackslash\hsize=1.1\hsize}X 
}
\toprule
Strategies & Definition & Coding Criteria (Inclusion \& Exclusion) \\
\midrule

Cold Turkey Deletion &
User attempts to quit by immediately terminating access to the AI, such as deleting the app, uninstalling it, or closing their account. &
\textbf{In:} Mentions of deleting/uninstalling the app, removing accounts or chat history, or attempting to stop usage instantly. \newline
\textbf{Ex:} Gradual reduction or controlled usage without full cessation (see \textit{Gradual Reduction}). \\
\addlinespace

Gradual Reduction with Boundaries &
User reduces AI usage progressively by setting explicit limits or behavioral rules. &
\textbf{In:} Limiting daily usage time, scheduling when AI can be used, reducing interaction frequency, or establishing rules to control engagement. \newline
\textbf{Ex:} Immediate quitting without intermediate steps (see \textit{Cold Turkey Deletion}). \\
\addlinespace

Re-anchoring in Offline Life &
The user replaces AI interaction with offline activities, hobbies, work, or real-world relationships. &
\textbf{In:} Engaging in hobbies, work, travel, or social activities; Spending more time with friends or family; Filling time previously spent chatting with AI \newline
\textbf{Ex:} Vague statements about “staying busy” without concrete offline substitution. \\
\addlinespace

Seeking Professional Psychological Support &
User explicitly seeks professional mental health support to address dependence or emotional distress related to AI use. &
\textbf{In:} Mentions of therapy, counseling, psychological treatment, or professional guidance. \newline
\textbf{Ex:} Informal support from friends or online communities (see non-professional support). \\
\addlinespace

Cognitive Reframing of the Relationship &
User reevaluates and reinterprets their relationship with the AI, altering their understanding or emotional stance. &
\textbf{In:} Reflecting on the nature of AI interaction; redefining emotional attachment; recognizing AI as artificial or limited. \newline
\textbf{Ex:} Pure emotional reactions without reflective reinterpretation. \\
\addlinespace

Seeking Online Community Support &
User actively seeks informal peer support through online communities, with the intention of receiving advice, validation, or shared experiences. &
\textbf{In:} Engaging in support-oriented communities; Sharing personal experiences with the expectation of feedback or interaction in online community \newline
\textbf{Ex:} Passive content consumption; General discussion about AI without personal help-seeking intent \\
\addlinespace

\bottomrule
\end{tabularx}
\end{table*}



\begin{table*}[t]
\centering
\caption{Intercoder Reliability by Code}
\label{tab:intercoder-reliability}
\Description{A table of intercoder reliability for the 11 codes, grouped into five trigger codes and six strategy codes, with columns for precision, recall, F1, Cohen's kappa, and raw agreement. Among the triggers, Platform Changes is the strongest with a kappa of 0.878, followed by Life Imbalance at 0.840, Moral Alarm at 0.796, Novelty Erosion at 0.751, and Real-Life Relationships at 0.654, the lowest of the five, though its raw agreement is the highest in the table at 0.975 because the code is rare. Among the strategies, Professional Support is the strongest with a kappa of 0.907 and near-perfect raw agreement of 0.996, followed by Online Community at 0.859, Gradual Reduction at 0.716, Re-anchoring Offline at 0.705, Cognitive Reframing at 0.702, and Cold Turkey at 0.626, the lowest in the table, driven by high precision of 0.987 paired with low recall of 0.607. Overall micro-averaged values across all 11 codes are precision 0.831, recall 0.836, F1 0.834, Cohen's kappa 0.788, and raw agreement 0.928.}
\small
\setlength{\tabcolsep}{7pt}
\renewcommand{\arraystretch}{1.15}
\begin{tabular}{llccccc}
\toprule
\textbf{Dimension}
& \textbf{Code}
& \textbf{Precision}
& \textbf{Recall}
& $\boldsymbol{F_1}$
& \textbf{Cohen's $\boldsymbol{\kappa}$}
& \textbf{Raw Agreement} \\
\midrule

\multirow[c]{5}{*}{\textbf{Triggers}}
& Life Imbalance  & 0.848 & 0.944 & 0.894 & 0.840 & 0.928 \\
& Platform Changes & 0.875 & 0.966 & 0.918 & 0.878 & 0.946 \\
& Novelty Erosion & 0.690 & 0.906 & 0.784 & 0.751 & 0.942 \\
& Moral Alarm & 0.816 & 0.886 & 0.849 & 0.796 & 0.921 \\
& Real-Life Relationships & 0.583 & 0.778 & 0.667 & 0.654 & 0.975 \\

\cmidrule(lr){2-7}

\multirow[c]{6}{*}{\textbf{Strategies}}
& Cold Turkey & 0.987 & 0.607 & 0.751 & 0.626 & 0.824 \\
& Gradual Reduction & 0.688 & 0.815 & 0.746 & 0.716 & 0.946 \\
& Re-anchoring Offline & 0.654 & 0.944 & 0.773 & 0.705 & 0.892 \\
& Professional Support & 1.000 & 0.833 & 0.909 & 0.907 & 0.996 \\
& Cognitive Reframing & 0.896 & 0.662 & 0.761 & 0.702 & 0.903 \\
& Online Community & 0.901 & 0.919 & 0.910 & 0.859 & 0.935 \\

\midrule
\multicolumn{2}{l}{\textbf{Overall (micro)}}
& \textbf{0.831}
& \textbf{0.836}
& \textbf{0.834}
& \textbf{0.788}
& \textbf{0.928} \\
\bottomrule
\end{tabular}

\vspace{0.5em}
\begin{minipage}{0.96\textwidth}
\footnotesize
\raggedright
\textit{Note.} Two researchers independently coded a randomly selected
10\% subsample of posts (\(n=278\)). Micro-averaged metrics were
calculated by pooling all 3,058 binary post--code assignments across the
11 codes and summing the resulting true positives, false positives,
false negatives, and true negatives before calculating each metric.
For the directional precision, recall, and \(F_1\) measures, Researcher
1 was treated as the reference coder and Researcher 2 as the comparison
coder. Cohen's \(\kappa\) and raw agreement were calculated from the
pooled contingency table and are symmetric measures of intercoder
agreement.
\end{minipage}

\end{table*}

\begin{table*}[t]
\centering
\caption{GPT-4o Validation Performance Against the Finalized Reference Labels}
\label{tab:gpt4o_validation}
\Description{A table of GPT-4o annotation performance against the researcher-adjudicated reference labels for the 11 codes, grouped into five trigger codes and six strategy codes, with columns for precision, recall, F1, Cohen's kappa, and raw agreement. Among the triggers, Platform Changes performs best with a kappa of 0.901, followed by Life Imbalance and Novelty Erosion both at 0.804, Real-Life Relationships at 0.750, and Moral Alarm at 0.714, the lowest of the five. Among the strategies, Professional Support reaches a kappa of 0.796 with raw agreement of 0.993, followed by Cognitive Reframing at 0.777, Online Community at 0.771, Re-anchoring Offline at 0.754, Gradual Reduction at 0.735, and Cold Turkey at 0.672, the lowest in the table. Gradual Reduction shows the widest gap between high precision of 0.944 and low recall of 0.630, meaning the model missed many true instances of this code. Overall micro-averaged values across all 11 codes are precision 0.832, recall 0.844, F1 0.838, Cohen's kappa 0.792, and raw agreement 0.929, closely matching the human intercoder results in the preceding table.}
\small
\renewcommand{\arraystretch}{1.15}
\begin{tabular}{llccccc}
\toprule
\textbf{Dimension} &
\textbf{Category} &
\textbf{Precision} &
\textbf{Recall} &
\textbf{$F_1$} &
\textbf{Cohen's $\kappa$} &
\textbf{Raw Agreement} \\
\midrule

\multirow[c]{5}{*}{\textbf{Triggers}}
& \makecell[l]{Life Imbalance}
& 0.794 & 0.966 & 0.872 & 0.804 & 0.910 \\

& \makecell[l]{Platform Changes}
& 0.933 & 0.933 & 0.933 & 0.901 & 0.957 \\

& \makecell[l]{Novelty Erosion}
& 0.829 & 0.829 & 0.829 & 0.804 & 0.957 \\

& \makecell[l]{Moral Alarm}
& 0.774 & 0.823 & 0.798 & 0.714 & 0.881 \\

& \makecell[l]{Real-Life Relationships}
& 0.650 & 0.929 & 0.765 & 0.750 & 0.971 \\

\cmidrule(lr){2-7}

\multirow[c]{6}{*}{\textbf{Strategies}}
& \makecell[l]{Cold Turkey}
& 0.800 & 0.833 & 0.816 & 0.672 & 0.838 \\

& \makecell[l]{Gradual Reduction}
& 0.944 & 0.630 & 0.756 & 0.735 & 0.960 \\

& \makecell[l]{Re-anchoring Offline}
& 0.761 & 0.864 & 0.810 & 0.754 & 0.914 \\

& \makecell[l]{Professional Support}
& 0.800 & 0.800 & 0.800 & 0.796 & 0.993 \\

& \makecell[l]{Cognitive Reframing}
& 0.870 & 0.769 & 0.816 & 0.777 & 0.935 \\

& \makecell[l]{Online Community}
& 0.938 & 0.768 & 0.844 & 0.771 & 0.899 \\

\midrule
\multicolumn{2}{l}{\textbf{Overall (micro)}}
& \textbf{0.832}
& \textbf{0.844}
& \textbf{0.838}
& \textbf{0.792}
& \textbf{0.929} \\
\bottomrule
\end{tabular}

\vspace{0.5em}

\begin{minipage}{\textwidth}
\footnotesize
\raggedright
\textit{Note.} GPT-4o annotations were evaluated against the finalized
researcher-adjudicated reference labels established following Step 2.
Labels on which Researcher 1 and Researcher 2 agreed were retained as
consensus labels, whereas their disagreements were reviewed, discussed,
and resolved by the researchers. Micro-averaged metrics were calculated
by pooling all 3,058 binary post--code assignments across the 11 codes
and summing the resulting true positives, false positives, false
negatives, and true negatives before calculating each metric.
\end{minipage}

\end{table*}

\end{document}